\documentclass[aps,prb,reprint,pdflatex,showpacs,superscriptaddress]{revtex4-2}
\usepackage{amsmath}
\usepackage{amssymb}
\usepackage{amsthm}
\usepackage{bbm}
\usepackage{graphicx}
\usepackage{hyperref}
\usepackage{physics}
\usepackage{systeme}
\usepackage{times}
\usepackage{epstopdf}
\usepackage{float}

\hypersetup{
  colorlinks=true,linkcolor=blue,citecolor=blue,
  filecolor=blue,urlcolor=blue,breaklinks=true
}

\begin{document}

\title{Quantum motion of a spinless particle in curved space:
  A viewpoint of scattering theory}

\author{Fabiano M. Andrade}
\email{fmandrade@uepg.br}
\affiliation{
        Departamento de Matem\'{a}tica e Estat\'{i}stica,
        Universidade Estadual de Ponta Grossa,\\
        84030-900, Ponta Grossa, Paran\'{a}, Brazil
      }
\author{Augusto R. Chumbes}
\email{august_rueda@hotmail.com}
\affiliation{
        Departamento de F\'{i}sica,
        Universidade Federal do Maranh\~{a}o,\\
        65085-580, S\~{a}o Lu\'{i}s, Maranh\~{a}o, Brazil
      }

\author{Cleverson Filgueiras}
\email{cleverson.filgueiras@dfi.ufla.br}
\affiliation{
        Departamento de F\'{i}sica,
        Universidade Federal de Lavras, Caixa Postal 3037,
        37200-000, Lavras, Minas Gerais, Brazil
      }
\affiliation{
  	Departamento de F\'{i}sica,
  	Universidade Federal da Para\'{i}ba, Caixa Postal 5008, 58051-900, Jo\~{a}o Pessoa, Para\'{i}ba, Brazil  }

\author{Edilberto O. Silva}
\email{edilberto.silva@ufma.br}
\affiliation{
        Departamento de F\'{i}sica,
        Universidade Federal do Maranh\~{a}o,\\
        65085-580, S\~{a}o Lu\'{i}s, Maranh\~{a}o, Brazil
      }

\date{\today }

\begin{abstract}
In this work, we study the scattering of a spinless charged particle constrained to move on a curved surface in the presence of the Aharonov-Bohm potential. We begin with the equations of motion for the surface and transverse dynamics previously obtained in the literature (Ferrari G. and Cuoghi G., Phys. Rev. Lett. \textbf{100}, 230403 (2008)) and describe the surface with non-trivial curvature in terms of linear defects such as dislocations and disclinations. Expressions for the modified phase shift, S--matrix and scattering amplitude are determined by applying a suitable boundary condition at the origin, which comes from the self-adjoint extension theory. We also discuss the presence of a bound state obtained from the pole of the S--matrix. Finally, we claim that the bound state, the additional scattering and the dependence of the scattering amplitude with energy are solely due to the curvature effects.
\end{abstract}
\pacs{03.65.-w, 03.65.Nk, 04.62.+v}

\maketitle

\section{Introduction}

The motion of a quantum particle constrained to move on a surface is a phenomenon that can be understood by the arising of forces that exist
only as a result of the surface geometry and the quantum mechanical
nature of the system.
The widely accepted formalism developed in this context is based on the simulation of the classical motion of a particle on a surface in quantum mechanics by forcing the particle to move between two parallel surfaces separated by a distance $d$ \cite{AoP.63.586.1971}. This formalism, known as thin-layer quantization, provides a result that has important physical implications in the description of the quantum mechanics of particles on surfaces. Namely, when the limit $d\rightarrow 0$ is established, one obtains an equation which differs from the usual Schr\"{o}dinger equation by an additional potential which depends on the curvature of the surface. Years later, in 1981, this idea was generalized by da Costa \cite{PRA.23.1982.1981}, who derived the Schr\"{o}dinger equation by
starting from the three dimensional one and then reducing it to a two-dimensional differential equation.
Following this procedure, he has shown that when a quantum point
particle moves confined to a surface embedded in ordinary three
dimensional Euclidean space, it is subjected to a geometric potential.
From his ideas, a more rigorous approach including the presence of an electric and magnetic field was proposed by Ferrari and Cuoghi
\cite{PRL.100.230403.2008}.
They have shown that there are no couplings between the fields and the
surface curvature.
Moreover, by making a proper choice of the gauge, the surface and
transverse dynamics are exactly separable.
Such a model was improved latter by considering the inclusion of the spin
of the particle by Wang \textit{et al.} \cite{PRA.90.042117.2014}.
Using the same thin-layer quantization scheme to constrain a quantum
particle on the surface together with a transformed spinor
representation, the authors have found the geometric potential and the
presence of an extra factor, which can generate additional spin
connection geometric potentials by the curvilinear coordinates
derivatives.

In a recent work, the thin-layer quantization procedure has been
refined and further developed by taking the proper terms of degree one
in $q_{3}$ ($q_{3}$ denotes the curvilinear coordinate variable
perpendicular to the curved surface) back into the surface quantum equation
\cite{AoP.364.68.2016}.
The thin-layer quantization formalism has been considered for a
variety of problems with different physical contexts (see Refs.  \cite{PRL.112.257203.2014,PRA.89.033630.2014,PRB.87.174413.2013,
PRA.88.033837.2013,PRE.88.043202.2013,PRX.4.011038.2014}).
It is also important to mention that other alternative approaches for
the confinement of a quantum particle on a surface are found in the
literature.
For instance, in Ref. \cite{PLA.380.1985.2016} a new formalism has been
proposed to construct the Hamiltonian of a spin-$1/2$ particle
with spin-orbit coupling confined to a surface that is embedded in a
three-dimensional space spanned by a general orthogonal curvilinear
coordinates.
In this approach, the authors consider a gauge field that allows us to
express the spin-orbit coupling as a non-Abelian SU(2) gauge field.
They also found that the geometric potential represents a coupling
between the transverse component of the gauge field and the mean
curvature of the surface that replaces the coupling between the
transverse momentum and the gauge field.
An extension of this approach was later accomplished in Ref.
\cite{PLA.380.2876.2016}.

In Ref. \cite{CMP.118.495.1988}, Deser and Jackiw studied the classical and
quantum scattering on a conical surface in (2+1) dimensions.
They considered a spinless charged quantum particle and an intrinsic
conical geometry for the system leading to a Hamiltonian which does not
include the geometric potential, $V_S(r)$.
Here, we consider the same system but with the addition of an
Aharonov-Bohm potential \cite{PR.115.485.1959} in the curved space that
includes the geometric potential, which arises from the da Costa
approach.
We show that our result for the scattering phase shift differs
from their result when we set $\phi=0$ (absence of AB flux).
This difference is solely due to the presence of the geometric
potential, showing an observable difference between the 2D quantum
mechanics on a conical surface and the 3D quantum mechanics
constrained to a 2D embedded conical surface.
Moreover, Deser and Jackiw do not found bound states.
The formation of bound states in curved or twisted surfaces was first
recognized by the works of Exner and Seba \cite{JMP.30.2574.1989} and
Goldstone and Jaffe \cite{PRB.45.14100.1992}.
The case of bound states in the present physical system was
considered by some of us in \cite{AoP.362.739.2015}, by using a
self-adjoint regularization procedure \cite{CMP.139.103.1991}.
Here, we derive the bound states energy from the poles of the scattering
matrix, confirming the results in our previous work and, once again,
proving that the presence of bound states in our system is also a
consequence of the geometric potential.

\section{Schr\"{o}dinger equation for a particle on a curved surface}

\begin{figure}
  \centering
  \includegraphics[width=\columnwidth]{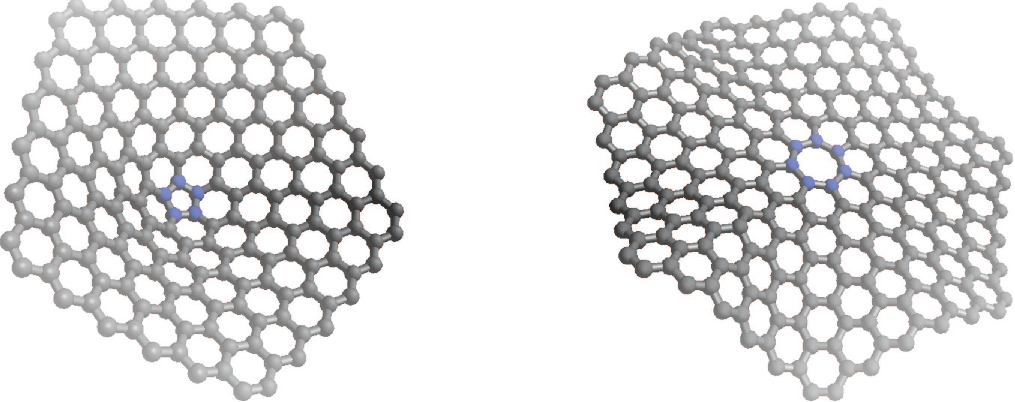}
  \caption{
    A conical surface made up by the inclusion of an one-pentagon
    apex ($\alpha<1$) (left) and a saddle-like surface made up by the
    inclusion of an one-heptagon apex ($\alpha>1$) on a honeycomb
    lattice (right).
  }
  \label{fig:defects}
\end{figure}

As mentioned above, we follow the da Costa's approach
\cite{PRA.23.1982.1981} in order to derive the general Schr\"odinger
equation for a quantum particle on a conical surface.
Before achieve this goal, we must mention again that a refinement of the
fundamental framework of the thin-layer quantization, considering the
surface thickness, was addressed in \cite{AoP.364.68.2016}.
Nevertheless, we will not consider such influence of the surface
thickness here since the respective extra terms can be treated using
perturbation theory.
Then, our case is valid for a quantum particle constrained to move in a
thin layer with constant width $d$, such that the constraint to the
curved surface is achieved in the limit $d\rightarrow 0$.
In what follows, the decomposition of the wave function is done under
this assumption.
This way, we begin by decomposing the total wave function $\psi$ into
its normal ($N$) and surface ($S$) components,
$\chi(q_{1},q_{2},q_{3},t)=\chi_{S}(q_1,q_{2},t) \chi_{N}(q_{3},t)$.
In this manner, the Schr\"{o}dinger equation can be decomposed into a
normal ($N$) and a surface ($S$) components as well
\cite{PRL.100.230403.2008}.
As a result, we have the normal component
$(\hbar=c=1)$,
\begin{equation}
  i\frac{\partial}{\partial t}\chi_{N}=
  \left[
    -\frac{\partial_{3}\partial^{3}}{2M}
    +V_{\lambda}(q_{3})
  \right] \chi_{N},
  \label{eq:normal}
\end{equation}
where $V_{\lambda}(q_{3})$ is the transverse potential, with $\lambda$
being the squeezing parameter \cite{PRA.23.1982.1981}, and the surface
components
\begin{align}
  i\frac{\partial}{\partial t}\chi_{S} = {}
  &
    \frac{1}{2M}\Bigg[
    -\frac{1}{\sqrt{g}}\partial_{a}
    \left(\sqrt{g}g^{ab}\partial_{b}\right)
  +\frac{iQ}{\sqrt{g}}\partial_{a}\left(\sqrt{g}g^{ab}A_{b}\right)
  \nonumber \\
   & +2iQg^{ab}A_{a}\partial_{b}+Q^{2}g^{ab}A_{a}A_{b}
    +V_{S}\Bigg] \chi_{S},
    \label{eq:surface}
\end{align}
where $g^{ab}$ is the contravariant component of the metric tensor of
the surface, $g=\det(g^{ab})$, $a,b=1,2$, $Q$ the charge of the particle,
$A_{j}$ the covariant components of the vector potential and
$V_{S}(q_{1},q_{2})$ is the potential due to the geometry of the surface.
As we are only interested in the dynamics on the surface,
Eq. \eqref{eq:normal} will be ignored in our approach.
On the other hand, Eq. \eqref{eq:surface} is just the two-dimensional
Schr\"{o}dinger equation for a spinless particle constrained to move on
$S$ by the normal potential $V_{\lambda}(q_{3})$.
We can see that  Eq. \eqref{eq:surface} includes the geometrical
potential $V_{S}(q_{1},q_{2})$ \cite{PRA.25.2893.1982}, which comes
from the two-dimensional confinement.
This potential is expressed in terms of the mean curvature
$\mathcal{H}$ and the Gaussian curvature $\mathcal{K}$  of the surface
\cite{PRA.23.1982.1981}:
\begin{equation}
  V_{S}(q_1,q_2)=-\frac{1}{2M}(\mathcal{H}^{2}-\mathcal{K}).
  \label{eq:ptg}
\end{equation}
By means of this geometric potential we study the physical implications of
the geometry on the dynamics of the particle.
In this work, we consider electrons constrained to move on a circular cone-shaped surface, which is described by the map
\begin{equation}
  X(\rho,\varphi)=
  \left(
    \rho\sin\left(\frac{\theta}{2}\right)\cos{\varphi},
    \rho\sin\left(\frac{\theta}{2}\right)\sin{\varphi},
    \rho\cos\left(\frac{\theta}{2}\right)
  \right),
\end{equation}
where $\rho$ is the distance along the cone from its apex, $\theta$ is
its apex angle and $0\leq\varphi\leq 2\pi$.
Such a map induces the metric
\begin{equation}
  \label{eq:metric}
  ds^2 = d\rho^2 + \alpha^2 \rho^2 d\varphi^2,
\end{equation}
where we have put $\alpha\equiv\sin(\theta/2)$.
In the geometric theory of defects, this parameter is the one
characterising the disclination on a solid \cite{AoP.216.1.1992}, which
is a kind of topological defect.
Our case deals with conical structures which appear in common
semiconductors \cite{JACS.127.13782.2005} and in graphitic carbon
materials as well \cite{PLA.342.237.2005,N.388.451.1997}.
In the former case, any value for $\alpha < 1$ can be engineered.
In the last one, to respect the symmetries of the carbon
network, we must have $\alpha=1-\lambda/2\pi$, with $\lambda=\pm
j\pi/3$, where $j$ is an integer in the interval $(0,6)$
\cite{N.388.451.1997}.
For example, if $j \equiv 1$, then $\alpha=1-\lambda/2\pi=5/6$
($\alpha=1+\lambda/2\pi=7/6$) stands for a graphitic sheet with positive
(negative) disclination where a single hexagon was substitute by an
one-pentagon (one-heptagon) apex, creating a conical (saddle-like)
structure, as can be seen in Fig. \ref{fig:defects} (see also
Ref. \cite{MPM.27.118.2016}).
Therefore, the metric \eqref{eq:metric} actually stands for two cases: a
conical surface when $0 < \alpha <1$ and a saddle-like surface when
$\alpha >1$ \cite{AoP.216.1.1992} (see Fig. \ref{fig:defects}).

Let us justify the usage of the da Costa's approach for a conical
surface.
Such a thin layer procedure was rigorously employed in
Ref. \cite{PRL.100.230403.2008} to derive the Schr\"{o}dinger equation
valid for \textit{any} 2D curved structure when magnetic and electric
fields are present.
Actually, it is valid for any orientable (not self-intersecting)
surface, but not for the non-orientable ones, such as the M\"{o}bius
strip or Klein bottle.
With a proper choice of the gauge, the dynamics on the surface is
analytically decoupled from the transverse one.
In Ref. \cite{PRL.100.230403.2008}, the case of a regular cylindrical
surface was examined.
In this case, a proper gauge was considered  which yields a magnetic
field which has a parallel and a perpendicular component to the surface
everywhere.
The choice of the gauge we consider in what follows poses no obstacle
towards such separation.
We mentioned the cylindrical surface for a reason: from the
\textit{projective geometry} \cite{Book.2003.Coxeter}, a cylindrical
surface is just a special case of a conical one.
It is as a limiting case of a cone whose apex is moved off to infinity
in a particular direction. At first, the da Costa's approach for a cone must be considered
excluding the conical tip.
This means that the Gaussian curvature is zero everywhere, as in the
case of a cylinder.
However, close to the apex of the cone, the Gaussian curvature is large
(although not infinite in the real case).
This effect has relevance for the discussion of scattering and bound
states.
We start by recalling that, by the Gauss-Bonnet
theorem \cite{Book.1977.Millman}, the conical tip yields the integral
over the conical surface of such Gaussian curvature different from
zero.
In fact, considering the area element on a cone given by
$dA=\sqrt{g}d\rho d\varphi$ with $g=\det(g_{ij})=\alpha \rho$, the
Gauss-Bonnet theorem yields \cite{PLA.342.237.2005},
\begin{equation}
   \label{eq:frank}
  \oint R\alpha \rho d\rho d\varphi =
  4\pi \left( 1-\alpha\right) = F,
\end{equation}
where $R$ is the Ricci scalar.
$F$ is the so called Frank vector, which gives the topological charge of
a conical surface (again $\alpha \equiv 1$ leads us to the flat space).
Then, a curvature flux exists and it is due to a short-range Gaussian
curvature, given by $\mathcal{K}={R}/{2}$.
The particle interacts with it.
However, this is a surface term, which means that it can be ignored in
the Schr\"odinger equation and treated via boundary conditions.
This is the procedure used in this work, where we consider it by taking
into account the self-adjoint extension approach
\cite{Book.2004.Albeverio}.
So, deriving the Schr\"{o}dinger equation for a cone is unnecessary,
since we are going just to repeat what has been made so far.

The Gaussian curvature, which is related to a geometric singularity, is
also  known as a conical singularity and was shown to reveal subtle
properties of quantum Hall states in \cite{PRL.117.266803.2016}.
The mean curvatures of the conical and saddle-like surfaces
(both are not self-intersecting surfaces) differ by a signal,
\begin{equation}
  \mathcal{H} = \frac{\sqrt{\pm (1-\alpha^{2})}}{2\alpha \rho},
  \label{eq:meancc}
\end{equation}
where here, and in what follows, we shall use the convention that the
upper sign is used for the conical surface and the lower sign is used
for the saddle-like surface.
This sign reversal occurs to be consistent with the fact that the
saddle-like surface has a negative curvature \cite{JMP.53.122106.2012}.
Therefore, the geometric potential $V_{S}(r)$ reads
\begin{equation}
  V_{S}(\rho)  = \frac{1}{2M}
  \left[
    \mp\frac{1-\alpha^{2}}{4\alpha^{2}\rho^{2}}
    +\mathcal{K}
 \right].
 \label{eq:vscone}
\end{equation}

We now introduce a magnetic flux tube parallel to the axis of the
surface.
In the background described by the metric \eqref{eq:metric}, the vector
potential, in the Coulomb gauge $\mathbf{\nabla}\cdot \mathbf{A}=0$, is
written as
\begin{equation}
  -QA_{i}= \epsilon_{ij}\frac{\phi r_{j}}{\alpha \rho^{2}}, \qquad
   A_{3}=0,
  \label{eq:vectorb}
\end{equation}
where $\epsilon_{ij}=-\epsilon_{ji}$ with $\epsilon_{12}=+1$,
$\phi =\Phi/\Phi_{0}$ is the flux parameter with $\Phi_{0}=2\pi /Q$.
In this manner, by writing the surface component of the wave function as
$\chi_{S}=e^{-iEt}\psi_{S}$, the Schr\"{o}dinger equation
\eqref{eq:surface}  results in
\begin{multline}
  -\frac{\partial^{2}\psi_{S}}{\partial \rho^{2}}
  -\frac{1}{\rho}\frac{\partial\psi_{S}}{\partial \rho}
  -\frac{1}{\alpha^{2}\rho^{2}}
  \left(\frac{\partial^{2}}{\partial\varphi^{2}}
    -\frac{2\phi}{i}\frac{\partial}{\partial \varphi}
  -\phi^{2}\right) \psi_{S}
  \nonumber \\
    \mp\frac{1-\alpha^{2}}{4\alpha^{2}\rho^{2}}\psi_{S}
    +\mathcal{K}\psi_{S}= k^{2}\psi_{S},
 \label{eq:schrof}
\end{multline}
where $k^{2} = 2 M E$.
We seek solutions of the form
\begin{equation}
\psi_{S}(\rho,\varphi) = e^{im\varphi} f_{m}(\rho),
\end{equation}
where $f_{m}(\rho)$ satisfies the eigenvalue equation
\begin{equation}
  h f_{m}(\rho) = k^{2} f_{m}(\rho),  \label{eq:diff}
\end{equation}
with
\begin{equation}
  h = h_{0}
  +\mathcal{K},
  \label{eq:grello}
\end{equation}
\begin{equation}
  h_{0}=
  -\frac{d^{2}}{d\rho^{2}}
  -\frac{1}{\rho}\frac{d}{d\rho}
  +\frac{j^{2}}{\rho^{2}},
  \label{eq:hzero}
\end{equation}
and
\begin{equation}
  j^{2} = \frac{4 (m+\phi)^{2}\mp (1-\alpha^{2})}{4 \alpha^{2}},
\label{eq:angular}
\end{equation}
is the effective angular momentum.
We can observe from Eq. \eqref{eq:angular} that some combinations of
the variables $m$, $\phi$ and $\alpha$ may lead to $j^{2}<0$, resulting
$j$ being a complex number \cite{AoP.362.739.2015}.
However, in this work, we shall only focus our
attention on $\alpha$ values that satisfy the condition
\begin{equation}
  \label{eq:constraint}
  4(m+\phi)^2 > \pm (1-\alpha^2).
\end{equation}

\section{The self-adjoint extension approach}

The presence of the singular Gaussian curvature, $\mathcal{K}$, in the
Hamiltonian $h$, as discussed above, turns it into a non-self-adjoint
operator.
The non-self-adjointness of $h$ stems from the fact that the
$\mathcal{K}$ must be singular at the cone apex.
Therefore, $h$ must be analyzed in the light of the self-adjoint
extension approach. The self-adjoint extension approach is a well-known
technique \cite{Book.2004.Albeverio,PRD.85.041701.2012,AoP.339.510.2013}.
However, to be self-contained, in this section we briefly review the
principal concepts and results of the self-adjoint extension approach.

Let us start by defining a self-adjoint operator.
A densely defined linear operator $\mathcal{O}$ on a Hilbert space is
said to be self-adjoint if it equals to its adjoint
$\mathcal{O}^{\dagger}$.
Then, two conditions must be fulfilled:
(i) the domain of $\mathcal{O}$ coincides with the domain of its
adjoint,
$\mathcal{D}(\mathcal{O})=\mathcal{D}(\mathcal{O}^{\dagger})$; and
(ii) the operator and its adjoint are equal in this domain,
$\mathcal{O}=\mathcal{O}^{\dagger}$.
As said above, $h$ is not self-adjoint.
However, it is natural to think $h$ as a self-adjoint extension of
$h_{0}$  given that fact that for smooth functions
$\xi \in C_{0}^{\infty}(\mathbb{R}^2)$ this two operators coincide at
the origin, $h \xi(0) = h_{0} \xi(0)$  \cite{JRAM.380.87.1987}.
Indeed, from the general theory of the self-adjoint extensions, it is
known that $h_{0}$ is self--adjoint only for $|j| \geq 1$, whereas for
$|j|<1$ it is not self-adjoint, has deficiency indices $(1,1)$ and
admits a one-parameter family of self-adjoint extensions
\cite{Book.1975.Reed.II}.
Then, we have to extend the domain of $h_{0}$ to its deficiency
subspace when $|j|<1$, which is spanned by the solutions of
$h_{0} f_{\pm}= \pm i k_{0}^{2}f_{\pm}$, where $k_{0}$ is a real parameter
introduced for dimensional reasons.
As showed in \cite{JMP.26.2520.1985}, all the self-adjoint extensions of
$h_{0}$ are parametrized by a boundary condition at the origin
\begin{equation}
\nu f_{0,j} = f_{1,j}, \label{eq:bc}
\end{equation}
where the boundary values are
\begin{align*}
  f_{0,j} = {} & \lim_{\rho\rightarrow 0^{+}}\rho^{| j|}f_{m}(\rho), \\
  f_{1,j} = {} & \lim_{\rho\rightarrow 0^{+}}\frac{1}{\rho^{| j|}}
                 \left[ f_{m}(\rho)-f_{0,j}\frac{1}{\rho^{|j|}}\right],
\end{align*}
and $\nu$ is the self-adjoint extension parameter defined in the
range $-\infty < \nu \leq \infty$.
The boundary condition in \eqref{eq:bc} describes $h_0$ plus a point
interaction at the origin (i.e., $h$).
The particular value $\nu=\infty$ is included in the range to represent
the free Hamiltonian, which corresponds to the case of the flat space,
$\alpha=1$, where the $\mathcal{K}$ strength vanishes and the Hamiltonian reduces to
\begin{equation}
  h_{\rm AB} =
  -\frac{d^{2}}{d\rho^{2}}
  -\frac{1}{\rho}\frac{d}{d\rho}
  +\frac{(m+\phi)^{2}}{\rho^{2}}.
\end{equation}
$h_{\rm AB}$ is the original AB Hamiltonian and, in this case,
only regular solutions contribute to the problem
\cite{PR.115.485.1959}.
On the other hand, it is not difficult to see that for $|\nu|<\infty$,
this boundary condition permits the contribution of irregular solutions
at the origin to the problem.

\section{Scattering and bound states analysis}

Now, we analyze the scattering and bound states for the
system represented by the Hamiltonian $h$.
We begin by writing the general solution of Eq. \eqref{eq:diff} for
$r\neq 0$,
\begin{equation}
  f_{m}(\rho) = a_{m} J_{|j|}(k\rho) + b_{m} J_{-|j|}(k\rho),
\label{eq:Bessel}
\end{equation}
where $J_{\upsilon}(z)$ is the Bessel function of fractional
order $\upsilon$.
The coefficients $a_{m}$ and $b_{m}$ represent the contributions of
the regular and irregular solutions at the origin, respectively.
As in the case of the original AB Hamiltonian discussed above, it is
very common in the literature discard the irregular solution
due to the normalizability condition.
However, due to the presence of the $\delta$ singularity in $h$, this cannot be done \cite{PRL.64.503.1990,PRD.50.7715.1994}.
We instead take into account both regular and irregular solutions and employ the boundary condition \eqref{eq:bc} to find out which irregular solutions are allowed by the self-adjoint extension.
Thus, by applying the boundary condition \eqref{eq:bc} to the general solution in \eqref{eq:Bessel}, it is possible to determine a relation between the coefficients $a_{m}$ and $b_{m}$.
From this relation, we can conclude that $b_{m}$ must be zero for
$|j|\geq 1$ and only the regular solution contributes to the wave
function in this case.
On the other hand, for $|j|<1$, we have the relation
\begin{equation}
  b_{m} = - \Omega_{j}^{\nu}(k) a_{m},
 \label{eq:relation}
\end{equation}
where
\begin{equation}
  \Omega_{j}^{\nu}(k)
  =
  \frac{
     k^{2| j|}
    \Gamma \left(1-|j|\right)
      \sin \left(|j| \pi \right)}
    { 4^{|j|}\Gamma \left(1+|j| \right) \nu
  + k^{2|j|}\Gamma \left(1-| j| \right)\cos \left(|j| \pi \right)},
 \label{eq:omega}
\end{equation}
and $\Gamma(z)$ represents the gamma function.
Therefore, the solution to the problem can be written as
\begin{equation}
  f_{m}(\rho) =
  \begin{cases}
    a_{m} J_{| j|}(k\rho), & \text{for } |j| \geq 1 \\
    a_{m}\left[J_{| j|}(k\rho)
    -\Omega_{j}^{\nu}(k) J_{-| j|}(k\rho)\right], &  \text{for } |j| < 1.
  \end{cases}
  \label{eq:super}
\end{equation}
Thus, we can see the contribution of the irregular solution for
the wave function is controlled by the self-adjoint extension
parameter.
The $\mathcal{K}$ induces a short-range potential, then it is possible to analyze it in terms of scattering phase shifts.
The phase shift, which measures how far the asymptotic scattering
solution of the problem differs from the asymptotic free solution,
can be obtained from the asymptotic expansion of the Bessel function in
Eq. \eqref{eq:super}, and the result seems
to be
\begin{equation}
  \delta_{j}^{\nu}\left(k\right) =
  \begin{cases}
    \delta_{j}^{\rm AB},
    & \text{for } |j| \geq 1\\
    \delta_{j}^{\rm AB}+\arctan\left[\Omega_{j}^{\nu}(k)\right],
    & \text{for } |j| < 1.
  \end{cases}
  \label{eq:phaseshift}
\end{equation}
where
\begin{equation}
  \label{eq:phaseshift_AB}
  \delta_{j}^{\rm AB}=\frac{\pi}{2}\left(|m| -|j| \right) ,
\end{equation}
is the AB scattering phase shift in the curved space.
It follows that the corresponding S--matrix (scattering matrix) is given by
\begin{align}
  S_{j}^{\nu}(k)
  = {}
  & e^{2 i \delta_{j}^{\nu}(k)}\nonumber\\
  = {}
  &
     \begin{cases}
    e^{2 i \delta_{m}^{\rm AB}},
    & \text{for } |j| \geq 1,\\
    e^{2 i \delta_{m}^{\rm AB}}
    \displaystyle
    \frac
    {1 + i \Omega_{j}^{\nu}(k)}
    {1 - i \Omega_{j}^{\nu}(k)}
    & \text{for } |j| < 1.
  \end{cases}
      \label{eq:smatrix}
\end{align}
Hence, the contribution of the irregular solution when $|j|<1$, which is controlled by the self-adjoint extension parameter $\nu$, causes a modification in the S--matrix when compared to the pure AB scattering
\cite{AoP.146.1.1983}.
Indeed, the result for the pure AB scattering is recovered if we set
$\nu=\infty$ or $\alpha=1$, the result is
$S_{m+\phi}^{\nu=\infty} = e^{2 i \delta_{m+\phi}^{\rm AB}}$  for all
$m$ values.

Another consequence of the contribution of the irregular solution is
the presence of bound states.
Bound states can be obtained from the poles of the S-- matrix in the upper half of the $k$ complex plane.
From Eq. \eqref{eq:smatrix}, it is easy to see that the poles come from $i \Omega_{j}^{\nu}(i \kappa)=1$, where we made the substitution
$k \to i \kappa$, $\kappa = \sqrt{-2 M E_{b}}$ and $E_{b}$
is the bound state energy.
Therefore, the explicit expression for the bound state energy is
\begin{equation}
  \label{bs_energy}
  E_{b} = -\frac{2}{M}
  \left[
    -\nu \frac{\Gamma(1+|j|)}{\Gamma(1-|j|)}
  \right]^\frac{1}{|j|},
\end{equation}
for $|j|<1$ and the pole occurs only for negative values of
$\nu$.
Energy (\ref{bs_energy}) is similar to the expression for the bound
state energy obtained in Ref. \cite{PLB.1991.268.222}, where the authors
studied the contact interactions of anyons.
We can also write down the scattering amplitude $f(k,\theta)$ in terms of the S--matrix:
\begin{align}
  f(k,\theta)
  = {}
  &
    \frac{1}{\sqrt{2\pi i k}}\sum_{m=-\infty}^{\infty}
    \left[
    S_{j}^{\nu}(k)-1
    \right]
    e^{i m \theta}   \nonumber \\
  = {}
  &
    \frac{1}{\sqrt{2\pi i k}}
    \Bigg\{
    \sum_{m\in \{|j| \geq 1\}}
    \left(e^{2 i \delta_{m}^{\rm AB}}-1\right)e^{im\theta}
    \nonumber \\
  &
    +
    \sum_{m\in \{|j| < 1\}}
    \left[
    e^{2 i \delta_{m}^{\rm AB}}
    \frac
    {1 + i\Omega_{j}^{\nu}(k)}
    {1 - i\Omega_{j}^{\nu}(k)}
    -1\right]
    e^{im\theta}
    \Bigg\}.
  \label{eq:amplitude}
\end{align}
We can observe that the scattering amplitude has a dependence on the
energy that goes beyond the usual energy scale which is, in the
non-relativistic domain, set by $1/k$ \cite{PRD.16.1815.1977}.
This additional energy dependence comes from the singularity which
arises from the geometric potential.
In other words, the singularity promoted by the localized curvature,
which is a conical topological defect, enhance the energy dependence of
the scattering amplitude.

\subsection{Discussion}

In Ref. \cite{CMP.118.495.1988}, the authors studied the classical and
quantum scattering on a conical surface in (2+1) dimensions.
They considered an intrinsic conical geometry leading to a Hamiltonian
which do not include the geometric potential, $V_S(r)$.
Their result for the scattering phase shift differs from our when we
set $\phi=0$.
This difference is solely due to the presence of the geometric
potential.
Moreover, they do not find bound states.
The presence of bound states in our system is also a consequence of the
geometric potential, which arises from da Costa's approach.
So here we can see an observable difference between the 2D quantum
mechanics on a conical surface and the 3D quantum mechanics
constrained to a 2D embedded conical surface.

In the light of the constraint in Eq. \eqref{eq:constraint}, in the case
of a conical surface, $\alpha<1$, the system does not have
bound states when $\phi=0$.
This is because of the effective angular momentum
is always outside the range $|j|<1$ for all values of $m$.
So, only the regular solution contributes to the problem.
On the other hand, for $\alpha>1$, there are values of $m$ for which
$|j|<1$, and there are bound states.

The fact that the scattering phase shift and the bound state energy
are dependent on the self-adjoint extension parameter could appear
strange at first.
However, it is possible to show by using another self-adjoint approach,
based on regularization of the $\delta$ potential
\cite{PRD.85.041701.2012,AoP.339.510.2013}, that the self-adjoint
extension parameter can be expressed in terms of physical parameters
\begin{equation}
  \nu =
  -\frac{1}{r_{0}^{2|j|}}
 \left(
    \frac
    {1 - \alpha + \alpha |j|}
    {1 - \alpha - \alpha |j|}
  \right),\label{pse}
\end{equation}
where $r_0$ is a very small radius which comes from the regularization
of the $\delta$ function.
This relation is valid only for $\nu<0$, because only for negative
values of the self-adjoint extension parameter we can have simultaneously
bound and scattering states which allow us to derive this relation.

\section{Conclusion}

We have thus explored the scattering scenario for the problem of a
spinless charged particle constrained to move on a curved surface,
with positive and negative curvatures, in the presence of the AB flux.
The particle is confined to move on the surface using the thin-layer
procedure proposed by da Costa \cite{PRA.23.1982.1981}.
The procedure gives rise to a geometric potential with $\delta$
function singularity at the origin, which is related to the tip of the
conical surface.
Due to this singularity, the Hamiltonian of the system is
not self-adjoint for all possible values of the effective angular
momentum $j$.
So, a suitable boundary condition, which comes from the theory of the
self-adjoint extensions was employed.
We have shown that this boundary condition allows the contribution of
irregular solutions at the origin for $|j|<1$.
As a result, the phase shift, S-matrix and scattering amplitude are
modified when compared with the pure AB scattering results.
In particular, due to the inclusion of irregular solutions, our
scattering results show additional energy dependence.
Moreover, from the poles of the S--matrix, an expression for the bound
state is obtained.
Finally, we have shown that the origin of this energy dependence is
solely due to the effects of curvature.
This comes from the geometric potential that arises from da Costa's
thin-layer approach.
Therefore, we can conclude that the additional dependence on the energy
of the scattering amplitude $f(k,\theta)$ is solely due to the effects of
localized curvature.

\acknowledgments
We thank the anonymous referees for valuable comments.
This study was financed in part by the Coordena\c{c}\~{a}o de
Aper\-fei\-\c{c}oamento de Pessoal de N\'{i}vel Superior - Brasil (CAPES) -
Finance Code 001, CNPq, FAPEMA, FAPEMIG and FAPPR.
FMA acknowledges CNPq Grants
313274/2017-7 and 434134/2018-0, and FAPPR Grant 09/2016.
EOS  acknowledges CNPq Grants 427214/2016-5 and 303774/2016-9, and FAPEMA
Grants 01852/14 and 01202/16.

\bibliographystyle{apsrev4-2}

\end{document}